\documentclass[runningheads]{llncs}
\usepackage{graphicx} 
\usepackage{xspace}
\usepackage{xcolor}
\usepackage{amsmath}
\usepackage{hyperref}
\usepackage[frozencache,cachedir=.]{minted}
\usepackage{newunicodechar}
\usepackage{tikz}
\usetikzlibrary{automata,positioning}
\usepackage{scalerel}
\usepackage{url}
\usepackage[only,llbracket,rrbracket,llparenthesis,rrparenthesis]{stmaryrd}
\usepackage{accsupp} 
\usetikzlibrary{cd}
\usepackage{algorithm}
\usepackage{stmaryrd}
\usetikzlibrary{tikzmark}
\usepackage{stix}
\tikzset{mycircled/.style={circle,draw,inner sep=0.01em,line width=0.01em}}
\newunicodechar{⊆}{\ensuremath{\subseteq}}
\newunicodechar{∃}{\ensuremath{\exists}}
\newunicodechar{↦}{\ensuremath{\mapsto}}
\newunicodechar{∧}{\ensuremath{\land}}
\newunicodechar{⇐}{\ensuremath{\Leftarrow}}
\newunicodechar{⬝}{\ensuremath{\bullet}}
\newunicodechar{↼}{\ensuremath{\leftharpoonup}}
\newunicodechar{Ξ}{\ensuremath{\mathbb{F}}}
\newunicodechar{β}{\ensuremath{\mathrm{\beta}}}
\newunicodechar{α}{\ensuremath{\alpha}}
\newunicodechar{⟧}{\ensuremath{\rrbracket}}
\newunicodechar{⟦}{\ensuremath{\llbracket}}
\newunicodechar{Π}{\ensuremath{\mathrm{\Pi}}}
\newunicodechar{∀}{{{$\forall$}}}
\newunicodechar{λ}{\ensuremath{\lambda}}
\newunicodechar{Φ}{\ensuremath{\mathrm{\Phi}}}
\newunicodechar{⊨}{\ensuremath{\vDash}}
\newunicodechar{⦃}{\ensuremath{\lBrace}}
\newunicodechar{⦄}{\ensuremath{\rBrace}}
\newunicodechar{∅}{\ensuremath{\emptyset}}
\newunicodechar{∪}{\ensuremath{\cup}}
\newunicodechar{∩}{\ensuremath{\cap}}
\newunicodechar{Δ}{\ensuremath{\mathrm{\Delta}}}
\newunicodechar{δ}{\ensuremath{\mathrm{\delta}}}
\newunicodechar{γ}{\ensuremath{\gamma}}
\newunicodechar{↔}{\ensuremath{\leftrightarrow}}
\newunicodechar{∈}{\ensuremath{\in}}
\newunicodechar{φ}{\ensuremath{\phi}}
\newunicodechar{⇈}{\ensuremath{\Uparrow}}
\newunicodechar{⊥}{\ensuremath{\bot}}
\newcommand{\conf}{\ensuremath{\mathfrak{c}}\xspace}
\newcommand{\toolName}{\textit{MMINT-A}}
\newcommand{\toolNamePL}{\textit{MMINT-PL}}
\usepackage{xspace}

\newcommand{\Conf}{\ensuremath{\elem{Conf}}}

\newcommand{\elem}[1]{\ensuremath{{\tt #1}}}
\newcommand{\token}[1]{{\small\elem{#1}}}

\newcommand{\process}{\ensuremath{\mathsf{PLACIDUS}}\xspace}
\begin{document}
\title{PLACIDUS: Engineering Product Lines of Rigorous Assurance Cases}
\author{
Logan Murphy\and
Torin Viger \and
Alessio Di Sandro \and 
Marsha Chechik}
\authorrunning{Murphy et al.}
%
\institute{University of Toronto, Canada\\
\email{\{lmurphy,tviger,adisandro,chechik\}@cs.toronto.edu}}

\vspace{-0.5in}
\maketitle
\begin{abstract}
\vspace{-0.3in}
In critical software engineering, structured assurance cases (ACs) are used to demonstrate how key properties (e.g., safety, security) are supported by evidence artifacts (e.g., test results, proofs). ACs can also be studied as formal objects in themselves, such that formal methods can be used to establish their correctness. Creating rigorous ACs is particularly challenging in the context of software product lines (SPLs), wherein a family of related software products is engineered simultaneously. Since creating individual ACs for each product is infeasible, AC development must be lifted to the level of product lines. In this work, we propose \process, a methodology for integrating formal methods and software product line engineering to develop provably correct ACs for SPLs. To provide rigorous foundations for \process, we define a variability-aware AC language and formalize its semantics using the proof assistant Lean. We provide tool support for \process as part of an Eclipse-based model management framework. Finally, we demonstrate the feasibility of \process by developing an AC for a product line of medical devices.
    
\end{abstract}
\vspace{-0.3in}
\section{Introduction}
{{In safety-critical software engineering, stakeholders require \emph{assurance} that software products will operate as intended. Several industries (e.g., automotive), have developed safety standards (e.g., ISO 26262 ~\cite{iso26262}) requiring careful documentation of verification activities via  \emph{assurance cases} (ACs)~\cite{rushby2015interpretation}. ACs use structured argumentation to refine system-level requirements into lower-level specifications which can be supported directly by evidence artifacts (e.g., tests, proofs). ACs can also been studied as formal objects in themselves, such that formal methods can be leveraged to verify their correctness~\cite{viger2023foremost,varadarajan2023clarissa}.}}
{Simultaneously, software systems are increasing in scale and complexity. In many cases, companies are not developing individual software products, but a {family} of related products, i.e., a {software product line} (SPL)~\cite{apel2016feature}. In such scenarios, sufficient assurance must be obtained for each product in the SPL. Ideally, this should be realized as the creation of a \emph{product line assurance case} (PL AC), from which product ACs can be {derived}.}

\begin{figure}[t]
    \centering
\includegraphics[width=\textwidth]{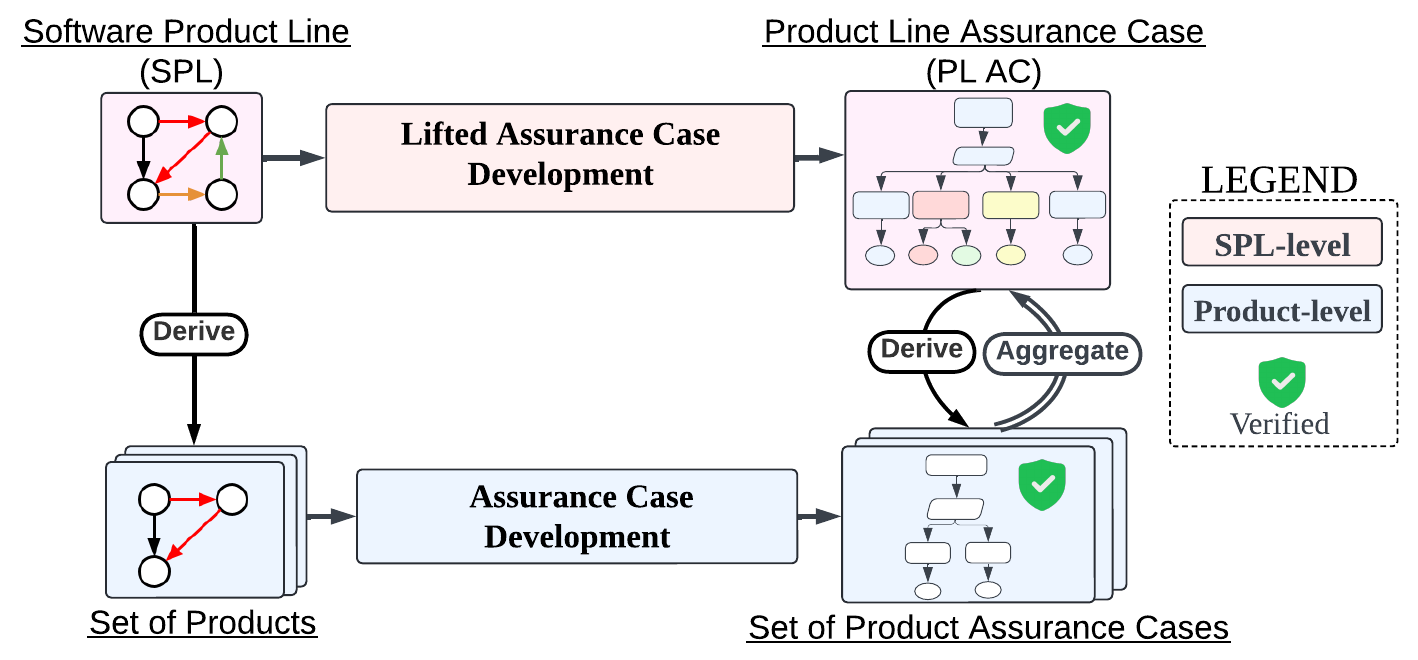}
    \vspace{-0.2in}
    \caption{Brute-force vs. lifted development of rigorous ACs for product lines.}
    \label{fig:liftvsbrute}
\end{figure}

Fig.~\ref{fig:liftvsbrute} illustrates two strategies for producing a verified PL AC. Beginning from an SPL (top left), the naive strategy is to derive the set of all products in the SPL (bottom left) and to develop verified product-level ACs for each product (bottom right) using existing techniques (e.g., ~\cite{viger2023foremost}). One can then aggregate the product-level ACs into a PL AC (top right) using some kind of variability encoding, such as the GSN patterns extension~\cite{habli2010safety}. This approach, which we refer to as the \emph{brute-force} method, is infeasible in practice, as many SPLs contain a very large number of products.

In SPL engineering (SPLE), \emph{lifting} is the process of redefining a product-level software analysis (e.g., model checking) so that it can be applied directly to product lines~\cite{Murphy2023ReusingYF}. The same idea can be applied to AC development: we need to \emph{lift} the AC development process (Fig.~\ref{fig:liftvsbrute}, top) such that we can take an SPL (top left) and produce a verified PL AC using SPL-level techniques (top right). However, there are two key roadbloacks to lifted AC development. 

First, there is more to AC development than software analyses: it crucially relies on \emph{argumentation}. While lifting software analyses is generally well-understood~\cite{Murphy2023ReusingYF}, it is less clear how to lift \emph{arguments}, how these two forms of lifting relate to one another, or how this lifting process should be integrated as part of a broader assurance engineering context. Second, although languages for PL ACs have been proposed in the past~\cite{habli2010safety}, the lack of specifically  variability semantics inhibits the rigorous lifting and analysis of PL ACs as formal objects. 

In this paper, we make the following contributions:
\vspace{-2mm}
\begin{enumerate}
    \item {We propose \process, an assurance engineering methodology which integrates formal methods and SPLE to support the lifted development of rigorous ACs.} 
    \item To provide rigorous foundations for \process, we formalize variational semantics of PL ACs using the proof assistant Lean~\cite{de2015lean}. Using this formalization, we show that verified PL ACs can be obtained by (i) lifting product-level argumentation structures and (ii) producing invariance proofs over variability-aware types.
    \item We provide tool support for \process as part of an Eclipse-based model management framework~\cite{mmint15}. 
\end{enumerate}
To demonstrate the feasibility of \process and the features supported by our tooling, we conducted a case study in which we developed a partial AC for a product line of medical devices. The rest of this paper is organized as follows. In Sec.~\ref{sec:background}, we provide the requisite background on AC development, and SPLE. In Sec.~\ref{sec:process}, we describe \process, our proposed methodology for lifted AC development. In Sec.~\ref{sec:gsnSPL}, we present rigorous formal foundations for \process, formalized in the proof assistant Lean. In Sec.~\ref{sec:tool}, we present our tooling and the case study.
In Sec.~\ref{sec:related}, we provide an overview of related work.
In Sec.~\ref{sec:conclusion}, we give concluding remarks and discuss future work.
\vspace{-0.1in}
\section{Background \& Related Work}
\label{sec:background}
 \vspace{-0.15in}
\subsection{Assurance Cases and GSN}
\label{sec:backgroundACS}
In this work, we focus on assurance cases represented using Goal Structuring Notation (GSN~\cite{kelly1999arguing}). 
{In GSN, an AC is modelled as a rooted tree, whose root and internal nodes are \emph{goals} (claims to be supported), and whose leaves are references to evidence artifacts, assumptions, or undeveloped goals. A goal can be {decomposed} into a finite set of subgoals. The resulting argument (referred to as a \emph{strategy}) is interpreted as a logical refinement: if each of the claims in the subgoals is true, then the parent goal should hold. If all strategies in the AC are sound refinements, we say the AC is \emph{deductive}. {An AC fragment for a hypothetical system is shown in Fig.~\ref{fig:gsn-example}. The root goal ${\tt G0}$ asserts a liveness claim about the system. A strategy over model checking (${\tt Str0}$) is performed, decomposing ${\tt G0}$ into four subgoals: that there is a representative behavioural model of the system $({\tt G1})$ and a specification which correctly formalizes the given property $({\tt G2})$, that model checking did not reveal any violations $({\tt G3})$, and that the verification procedure itself is actually sound $({\tt G4})$. Each of these subgoals requires either evidence (solution nodes, e.g. ${\tt Sn.1}$) or further decomposition. In Fig.~\ref{fig:gsn-example}, goal ${\tt G1}$ is left undeveloped.}}

\begin{figure}[t]
    \centering
\includegraphics[width=0.95\textwidth]{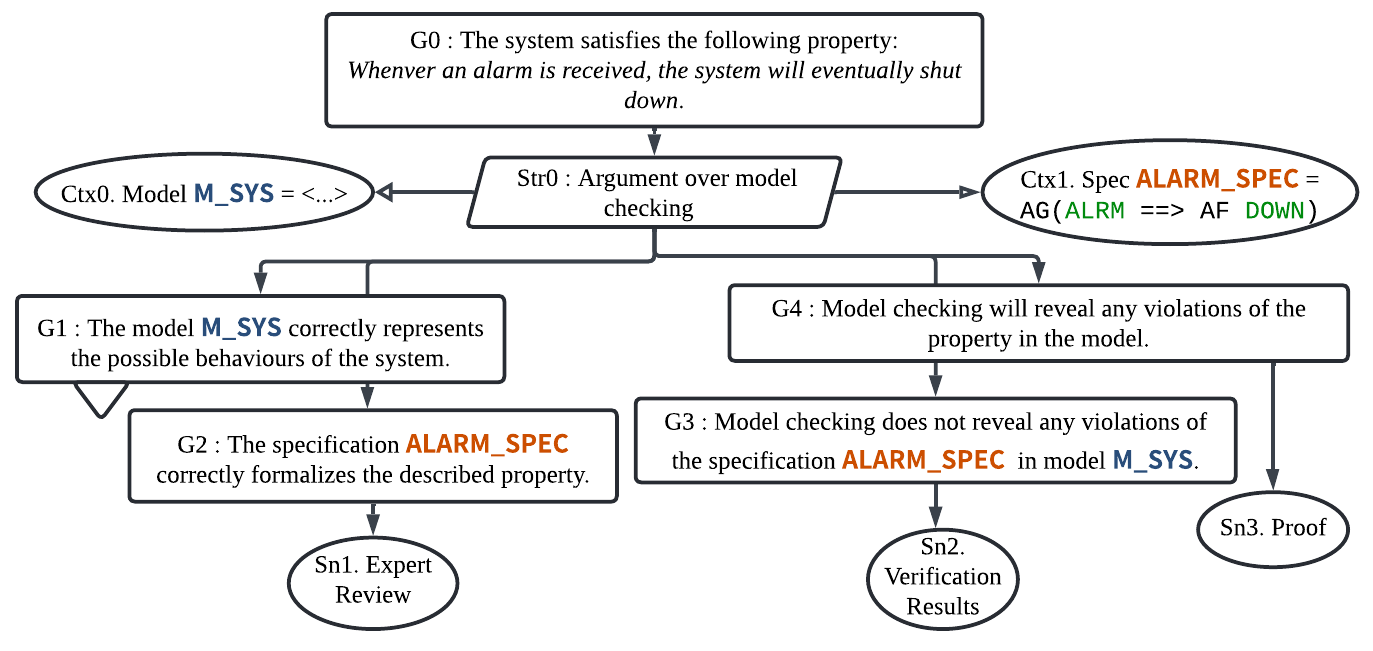}
\vspace{-0.1in}
    \caption{An AC fragment (in GSN) for a hypothetical system. This AC is obtained by instantiating a generic model checking template with model ${\tt M\_SYS}$ and CTL specification ${\tt ALARM\_SPEC}$.}
    \label{fig:gsn-example}
    \vspace{-0.1in}
\end{figure}
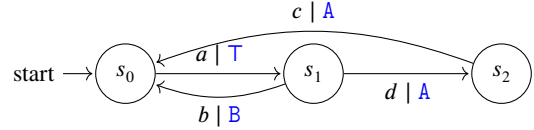
\begin{figure}[t]
\vspace{-0.2in}
    \centering
    \[\begin{tikzpicture}[shorten >=1pt,node distance=2.5cm,on grid,auto] 
   \node[state,initial] (q_0)   {$s_0$}; 
   \node[state] (q_1) [right=of q_0] {$s_1$}; 
   \node[state] (q_2) [right=of q_1] {$s_2$}; 
    \path[->] 
    (q_0) edge  node {$a \mid \textcolor{blue}{{\top}}$} (q_1)
    (q_2) edge [bend right=20, above] node {$c \mid \textcolor{blue}{{\tt A}}$} (q_0)
    (q_1) edge [below] node  {$d \mid \textcolor{blue}{{\tt A}}$}(q_2)
          edge [bend left=20, below] node {$b \mid \textcolor{blue}{{\tt B}}$} (q_0);
\end{tikzpicture}\]
\vspace{-9mm}
    \caption{A Featured Transition System (FTS) over features $\mathbb{F} = \{{\tt A}, {\tt B}\}$ and feature model $\Phi = {\tt A} ~{\tt xor}~ {\tt B}$ (adapted from ~\cite{beek2019static}).}
    \label{fig:FTS}
    \vspace{-0.25in}
\end{figure}

Ideally, deductive correctness of ACs is verified by formalizing its arguments and verifying the refinements, but this is a costly process. One method for facilitating the creation of sound arguments is the use of \emph{templates}~\cite{viger2020just}. {An argument template is a method for decomposing a goal into a set of subgoals using some predefined construction.} We say that a template is \emph{valid} if each argument which can be instantiated from the template is sound. {For example, the argument shown in Fig.~\ref{fig:gsn-example} is an instantiation of a (valid) argument template for model checking. Instantiating this particular template requires providing the model and specification used for verification (i.e., ${\tt M\_SYS}$ and ${\tt ALARM\_SPEC}$ in Fig.~\ref{fig:gsn-example}).}

\vspace{-0.15in}
\subsection{Software Product Lines}
{A \emph{software product line} (SPL) is a family of software artifacts ({products}) with distinct (but often overlapping) structure and behaviours~\cite{apel2016feature}. The variability of a  SPL is defined in terms of a set $\mathbb{F}$ of \emph{features}, each of which can be either present or absent in a given product. {A product is obtained from the SPL  by choosing a \emph{configuration} $\conf \subseteq \mathbb{F}$ of features.} Variability in an SPL can be modelled via \emph{feature expressions,}, i.e., propositional expressions whose atomic propositions are (boolean) feature variables. This induces a natural entailment relation $\conf \vDash \phi$ between feature expressions and configurations. Given a feature expression $\phi$ and configuration $\conf$, we use the notation $\Conf(\phi)$ to denote the the set of configurations satisfying $\phi$. The set of products which are \emph{valid} in an SPL is defined via a \emph{feature model} $\Phi$, also typically given as a feature expression.

{Beyond the features and feature model, an SPL must specify a set of domain assets from which products can be configured. In the context of \emph{annotative} product lines, the SPL is given as a collection of domain elements {annotated} with feature expressions, referred to as \emph{presence conditions}. For example, a product line of transition systems can be represented as a Featured Transition System (FTS,~\cite{classen2010model}).
{Fig.~\ref{fig:FTS} illustrates an FTS  (adapted from \cite{beek2019static}) over the feature set $\{{\tt A}, {\tt B}\}$ with feature model $\Phi = A~{\tt xor}~B$. The presence conditions of transitions are written in blue font next to the transition labels. There are two valid configurations of this SPL, as $\Conf(\Phi) = \{\{{\tt A}\},\{{\tt B}\}\}$. A product transition system is \emph{derived} under configuration $\conf \in \Conf(\Phi)$ by removing transitions whose presence conditions are not satisfied by $\conf$. In Fig.~\ref{fig:FTS}, the transition labelled by $a$ is present under both configurations, since it is annotated by $\top$, while the transition labelled by $b$ is present only in configuration $\{{\tt B}\}$. Similar derivation operators can be defined for other types of annotative product lines (e.g., product lines of code, trees, or sets). Given any annotative SPL $x$ and any configuration $\conf$, we can use the notation $x|_\conf$ to denote the product derived from $x$ under $\conf$.} {A key problem in SPLE is \emph{lifitng}, i.e., the redefinition of an analysis such that it can be applied to SPLs.}
\vspace{-2mm}
\begin{definition}[Lifting~\cite{Murphy2023ReusingYF}]\label{def:lift} \rm
    Let $A$ and $B$ be types, and let $\alpha$ (resp. $\beta$) be the type of ``product lines of $A$'' (resp. $B$). Let $f : A \to B$ be some function. Then a \emph{lift} of $f$ is a function $F : \alpha \to \beta$ such that for all feature models $\Phi$, all $x \in \alpha$ and all $\conf \in \Conf(\Phi)$, we have $F(x)|_\conf = f(x|_\conf)$.
\end{definition}

In some cases, the correctness criterion for lifting can be relaxed if the ``lifted'' analysis is at least \emph{sound} with respect to the original analysis. Consider, for instance, a hypothetical verifier $\mathsf{V}$ which performs model checking on FTSs. Let $M$ be some FTS we wish to verify against property $\phi$, and suppose that there are several products derivable from $M$ with \emph{distinct} counterexamples to $\phi$. In order for $\mathsf{V}$ to be a lift in the strict sense of Def.~\ref{def:lift}, $\mathsf{V}$ would need to report each of these counterexamples and their respective configurations of $M$. {SNIP}~\cite{classen2012model} is an example of a lifted model checker which performs this kind of analysis. However, there are other SPL-level model checkers (e.g., FTS2VMC~\cite{ter2022efficient}) which only report a single counterexample if one exists for \emph{some} product. Such tools are still {sound} in the sense that a successful verification of the SPL means that all products are verified. We refer to SPL-level analyses which are sound in this sense, but which do not satisfy Def.~\ref{def:lift}, as \emph{quasi-lifts}.

\section{Building Product Lines of Assurance Cases with PLACIDUS}
\label{sec:process}
We now describe \process\footnote{\textbf{P}roduct \textbf{L}ine \textbf{A}ssurance \textbf{C}ases v\textbf{I}a \textbf{D}ed\textbf{U}ction and analy\textbf{S}is}, a methodology for lifted AC development. \process begins from the observation that supporting SPL-level AC development requires {lifting} the tools used by AC developers to create rigorous product ACs. \process focuses on two kinds of development tools: AC templates (for structured argumentation) and software analyses (for evidence production). To maintain the rigor of the assurance process, this lifting must be approached {formally}. As such, \process is an interdisciplinary process which integrates expertise in assurance engineering, formal methods and SPLE. {We present \process as an incremental extension of two simpler development methodologies, as illustrated in Fig.~\ref{fig:workflows}.}


\begin{figure}[t]
    \centering
    \includegraphics[width=\textwidth]{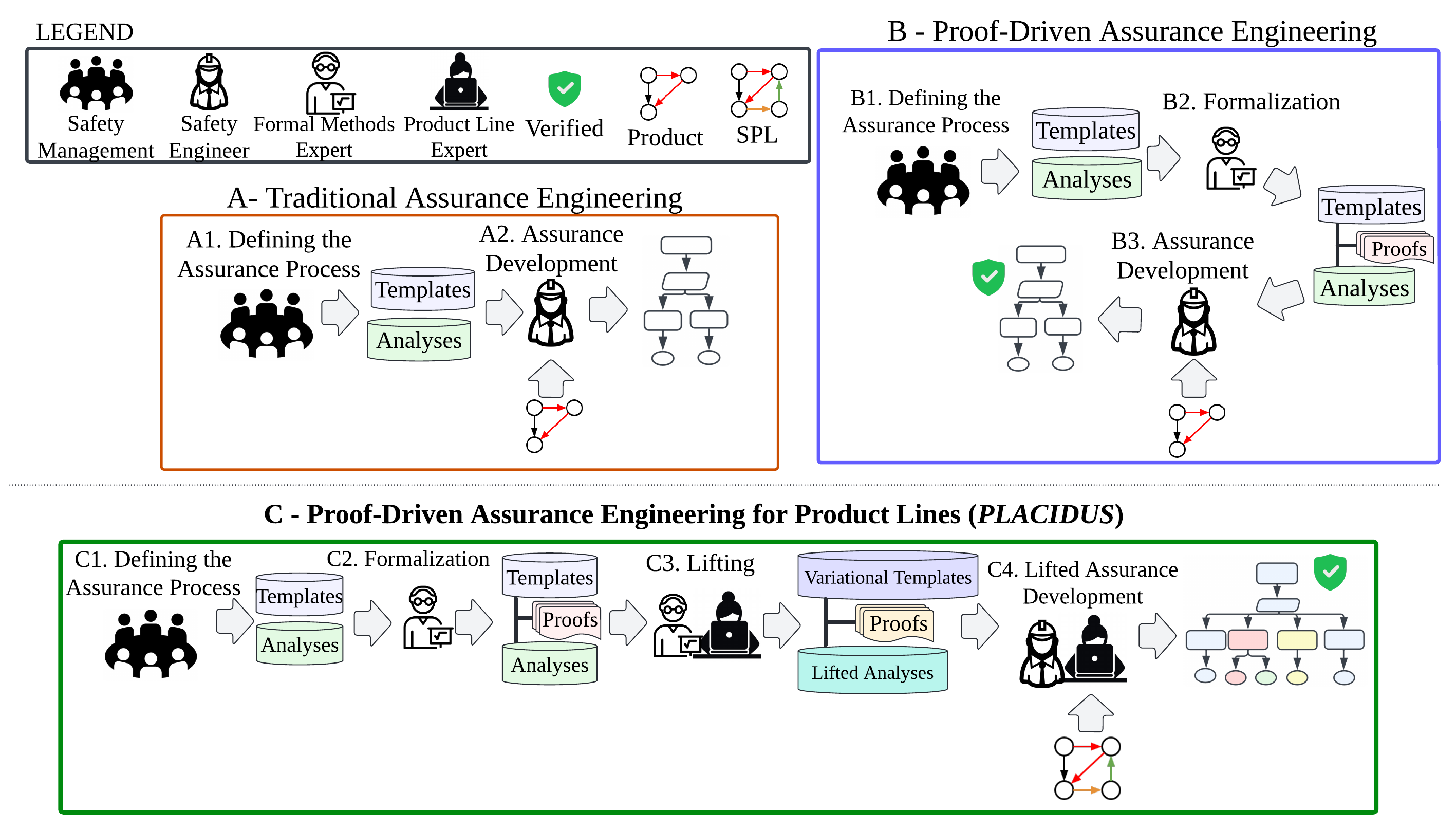}
    \vspace{-0.2in}
    \caption{Three assurance engineering methodologies: traditional assurance engineering (A); proof-driven assurance engineering (B); lifted, proof-driven assurance engineering via \process (C). 
    }
    \label{fig:workflows}
    \vspace{-0.2in}
\end{figure}


\noindent\textbf{Traditional Assurance Engineering.} {While there is no single ``canonical'' assurance engineering methodology, industry standards such as ISO 262626 outline general methodological workflows, of which the ``Traditional Assurance Engineering'' methodology (labelled A in Fig.~\ref{fig:workflows}) is a very simple instance.} In this workflow, there are two parties contributing to AC development: safety managers and safety engineers. Before any concrete assurance work is performed for a software product, the safety management team must define the assurance {process} which is to be followed by the safety engineers (A1). The assurance process is informed by various factors, such as industry standards (e.g., ISO 26262), corporate procedures, and available resources. The process definition includes, in particular, the forms of argumentation (templates) and the analyses engineers can use to develop the AC. The safety engineers can develop an AC for the software product under study by instantiating templates to form structured assurance arguments and running analyses to collect evidence artifacts. The end result is an (informal) AC for the software product.

\noindent\textbf{Proof-driven Assurance Engineering.}
One of the limitations of the traditional methodology is that the correctness of purely informal ACs needs to be validated manually, which is difficult, expensive and error-prone. If we instead impose formalization as part of AC development, we can use formal methods to verify the correctness of ACs. We refer to this methodology as \emph{proof-driven assurance engineering} (labelled B in Fig.~\ref{fig:workflows}). This has been an active research area in assurance engineering, with various formal methods being leveraged to improve the rigor of ACs, including automated theorem provers~\cite{denney2012advocate}, model checking~\cite{sljivo2018tool} and proof assistants~\cite{foster2021integration,viger2023foremost}. Like the traditional methodology, the first stage defines the assurance process (B1). Unlike the traditional methodology, a formal methods expert is engaged to \emph{formalize} the assurance process (or parts thereof) (B2). While there are variations in the research literature, most proposals use some version of the following procedure: (i) the semantics of relevant argument structures (templates) is formalized in an appropriate logic; (ii) based on this formalization, argument-level correctness proofs are given; (iii) analyses are provided to produce proofs as assurance evidence. Once the necessary formalization, proofs, and analyses have been developed (B2), the AC developer can  use the formalized argument structures and associated proofs to produce a verified (or partially verified) assurance case for the software product under study (B3).

\vspace{2mm}
\noindent\textbf{PLACIDUS.} We now present \process, which is our extension of the proof-driven methodology (B) to support lifted AC development (as per Fig.~\ref{fig:liftvsbrute}). The first two stages of \process (C1, C2) are exactly the same as in the proof-driven methodology (B1, B2), producing a set of formalized argument templates, their correctness proofs, and associated analyses. The core of \process is the \emph{lifting} stage (C3) which formally lifts these components to the SPL-level. More concretely: (i) analyses used as part of product-level AC development must be lifted; and (ii) formalized product-level templates must be lifted such that they can be instantiated with SPL-level data. The lifting of software analyses, either as strict lifts or quasi-lifts, is a very well-studied area of SPLE~\cite{thum2014classification,Murphy2023ReusingYF}, with a wide variety of common analyses having been lifted by researchers. But to the best of our knowledge, the only instance of lifted assurance argument template was proposed by Nešić et al.~\cite{nevsic2021product} for a contract-based argumentation template. In \process, we expect this lifting to be performed for \emph{all} argument templates. As with the product-level formalization (C2), part of the outcome of C3 is a set of correctness proofs (or proof-producing analyses) associated with the variational templates. But since we are now performing all analysis and reasoning at the SPL-level, the proofs themselves need to become ``{variational}''. 

As illustrated in Fig.~\ref{fig:workflows}, we expect the lifting stage (C3) to require collaboration between experts in formal methods and SPLE. In the lifted development stage (C4), safety engineers will have the verified variational templates and associated analyses at their disposal for AC development. We expect that the lifted development process will also require collaboration between safety engineers and SPLE experts, since SPL-level data will need to be interpreted as part of AC development. Since all AC development tools (templates, analyses) are now variational, they can be applied directly to the SPL under study, avoiding redundant product-level work\footnote{Of course, there may be some product-level work which cannot be avoided; strictly speaking, \process seeks to eliminate \emph{unnecessary} product-based work.}. The outcome of stage C4 is a \emph{product line of verified ACs}, such that a verified AC can be derived for every valid product in the SPL. Assuming such an object is well-defined, \process supports the lifted AC development as outlined in Fig.~\ref{fig:liftvsbrute}, although it remains to be seen what SPL-level semantics should be used for PL ACs, and what it means to lift and verify AC templates for SPLs.

\vspace{-0.1in}
\section{Formal Foundations of Lifted Assurance Case Development}
\vspace{-4mm}

\label{sec:gsnSPL}
In Sec.~\ref{sec:process}, we described a methodology for integrating formal methods and SPLE to support the lifted development of PL ACs. To implement this methodology, rigorous foundations are required. To this end, we have formalized a theory of PL AC semantics using the proof assistant Lean~\cite{de2015lean}.  In addition to formalizing the semantics of PL ACs, we are able to formalize the {lifting} stage of \process (C3 in Fig.~\ref{fig:workflows}). In particular, we show that lifting and verifying AC templates for SPLs reduces to (i) lifting functions viz. Def.~\ref{def:lift}, and (ii) creating invariance proofs over sets of products. Due to space limitations, we give a high-level (and slightly simplified) overview of our formalization. Further details are given in Appendix~\ref{app:formal}; the complete formalization is available online.
\footnote{\url{https://github.com/loganrjmurphy/placs}.}

\textbf{Product ACs and Templates.}
We first define a Lean type \token{GSN} as a minimal GSN-like language of ACs, comprising goals, decomposition strategies and evidence. Our language follows the GSN grammar given by Matsuno et al.~\cite{matsuno2014design}. The type \token{Goal} is a wrapper for propositions of the form $P(x)$, where $P$ is a predicate over some type $A$ and $x \in A$. Evidence for goal $\token{g}$ is given by providing a proof of the proposition represented by $\token{g}$. Strategies are tuples of the form $(\token{g}, \{\token{A}_1,...\token{A}_n\})$, where \token{g} is the parent node and $\token{A}_i$ are the children (i.e., \token{GSN} subtrees). Since \token{GSN} is a Lean type like any other, we can prove theorems about its terms. For instance, we can define the predicate \token{refines} asserting that a strategy $(\token{g}, \{\token{A}_1,...\token{A}_n\})$ satisfies $\left(\bigwedge_i \token{g}_i \right)  \Rightarrow \token{g}$, where $\token{g}_i$ is the root goal of $\token{A}_i$. We then define a recursive predicate \token{deductive} asserting that every strategy in the AC satisfies \token{refines.} Using these predicates, we can formally verify the deductive correctness of ACs formalized in our \token{GSN} language.

We next formalize GSN templates. Intuitively, argument templates are functions which take some existing AC claim as input, and produce a list of sub-ACs as output. In most cases, some auxiliary data is required to perform the instantiation -- for instance, decomposing goal \token{G0} in Fig.~\ref{fig:gsn-example} using a model checking template required providing the model \token{M\_SYS} and specification \token{ALARM\_SPEC} as inputs to the template. In our Lean formalization, we define a record type \token{Template} which includes the predicate \token{parent} of the goal being decomposed, and its instantiation function. Instantiation functions have the form $\token{inst} : A \times D \to \token{List}~\token{GSN}$, where $A$ is the domain of the claim being decomposed and $D$ is some type of auxiliary data. We then formalize a predicate \token{valid} which asserts that every instantiation of a template results in a sound argument. In practice, many templates are only \emph{conditionally} valid, i.e.,  the resulting argument is sound only if the instantiation data $(a,d) : A \times D$ satisfies some precondition. For example, consider using the \emph{domain decomposition} template given by Viger et. al~\cite{viger2020just} to decompose an {invariance} claim, i.e., a goal of the form $\forall x \in S.\: P(x)$ where $S$ is some set over type $A$. The template is instantiated for $S$ by giving a finite family $\mathcal{F} = \{X_1,..X_n\}$ of subsets of $A$, creating $n$ subgoals $\{g_1,..g_n\}$ such that goal $g_i$ asserts $\forall x \in X_i.\: P(x)$. The resulting argument is deductive if the family $\mathcal{F}$ is \emph{complete} w.r.t. $S$, i.e. $\bigcup_i X_i \subseteq S$. We can formalize this class of templates in Lean, and prove its (conditional) validity. This process -- formalizing and verifying templates -- corresponds to step C2 of \process (Fig.~\ref{fig:workflows}).

\textbf{Variational Types and Proofs.} We next formalize \emph{variational types}, which are an algebraic generalization of  product lines~\cite{Murphy2023ReusingYF} based on derivation operators. To formalize variational types in Lean, we must first formalize features, configurations, feature expressions, and the semantics of feature expressions. Presence conditions (\token{PC}) and feature models (\token{FeatModel}) are defined as type aliases for feature expressions. 

We say that  $\alpha$ is a \emph{variational type} if there exists a derivation function from $\alpha$ to some (product-level) type $A$. In Lean, this is captured using a typeclass ${\tt Var}$. We refer to such a type $\alpha$ as a \emph{variational extension} of the type $A$. Intuitively, to say that $\alpha$ is a variational extension of $A$ means that $\alpha$ is the type of ``product lines of $A$''. In Lean, we implement variational extensions using a typeclass $\token{Var}$:

\vspace{-2mm}
\begin{minted}[fontsize=\footnotesize]{lean}
class Var (α : Type) (A : Type) where
  derive : α → Conf Φ → A
\end{minted}
\vspace{-2mm}
where $\tt Conf~Φ$  is the type of valid configurations under feature model $\Phi$.
In what follows, we use Roman letters ($A,B$, etc.) to denote product-level types, and Greek letters $(\alpha, \beta$, etc.) to denote variational types. In Lean, we use the syntax ``$\token x \downarrow \conf$'' to denote derivation of $x$ under configuration $\conf$ for any instance of ${\tt Var}$ (equivalent to the notation $x|_\conf$). We can then define basic instances of the $\token{Var}$ typeclass, e.g., lists of elements annotated by presence conditions ($\token{List}~(A \times \token{PC})$) form a variational extension of ($\token{List}~A$), with derivation under configuration $\conf$ defined as filtering out elements whose presence conditions are not satisfied by $\conf$\footnote{Moreover, we can create a variational extension of any base type $T$ using the variability-aware type constructor defined by Shahin et al.~\cite{shahin2020automatic}.}.

As we will demonstrate shortly, reasoning about the correctness of PL ACs reduces to reasoning about a specific class of proofs -- namely, proofs of product-level invariants over a given set of products (e.g., the products of an SPL).
\vspace{-0.2mm}
\begin{definition}[Variational Proof] \rm Let $\tau$ be a variational extension of type $T$; let $P$ be a predicate over $T$; let $x \in \tau$, and let $\Phi$ be a feature model. Then a proof of $\forall \conf \in \Conf(\Phi).\:P(x|_\conf)$ is referred to as a \emph{variational proof}.
\end{definition}
\vspace{-0.2mm}
We may not always need a proof for every configuration $\conf \in \Conf(\Phi)$, but only a subset, i.e., $\Conf(\phi) \subseteq \Conf(\Phi)$, for some feature expression $\phi$. In Lean, we use $[\Phi]~\token{P}~\token{x}$ to denote the type of an (unrestricted) variational proofs, and $[\Phi|~\phi]~\token{P}~\token{x}$ for a proof restricted to $\Conf(\phi)$.

\textbf{Variational GSN.}
We now have the components needed to define a variational extension of the type \token{GSN}, which we call \token{vGSN}. Following Habli and Kelly~\cite{habli2010safety}, our language for PL ACs supports both \emph{structural variability} (i.e., certain fragments of the AC are only relevant for certain products) and \emph{semantic variability} (i.e., the meaning of a claim depends on the choice of feature configuration).  We begin by defining the type \token{vGoal} of variational goals, which is parameterized by a feature model $\Phi$. To model {structural} variability, each \token{vGoal} carries a presence condition, as proposed by Shahin~\cite{shahin2021towards}. To model semantic variability, we define (predicative) variational goals to consist of {product}-level predicates and variational data (i.e., a term of a variational type). It follows that the interpretation of a variational goal is no longer a truth-value, but a {function} from configurations to truth-values. We model evidence in \token{vGSN} using variational proofs: given a variational goal defined by predicate $P$, variational data $x$ and annotated by presence condition $\phi$, evidence for this goal is a term of type $[\Phi|~\phi]~\token{P}~\token{x}$. The inductive definition of \token{vGSN} is as follows:

\vspace{-2mm}
\begin{minted}[fontsize=\footnotesize]{lean}
inductive vGSN (Φ : FeatModel Ξ) 
| evd [Var α A] (φ : PC Ξ) (P : A → Prop) (x : α) (e: [Φ| φ] P x)
| strategy (g : vGoal Φ) (As : List (vGSN Φ))
\end{minted}
\vspace{-2mm}

We next define a derivation operator from \token{vGSN} to \token{GSN}. Given a configuration $\conf$, we recursively descend through the PL AC, checking whether each node's presence condition is satisfied by ${\conf}$. If the presence condition is not satisfied, the node is not present under $\conf$, so we return the empty AC \token{nil}. Otherwise, if we are at an evidence node, we derive the product-level proof of $\token{P}(\token{x}|_\conf)$ from the variational proof $\token{e}$. If we are at a strategy node, we derive the parent goal \token{g}$|_\conf$ and map the derivation operator over the children, removing instances of \token{nil}. This derivation operator allows us to register $\token{vGSN}$ as a variational extension of \token{GSN} in Lean.

In addition to their use as evidence for variational goals, variational proofs are also used to demonstrate correctness of the AC itself. For instance, recall the criterion \token{refines} which asserts that a decomposition strategy is logically sound. To prove that a {variational} strategy is sound, we need to provide a variational proof as follows: 
\vspace{-2mm}
\begin{minted}[fontsize=\footnotesize]{lean}
def refines (g : vGoal Φ) (l : List (vGSN Φ)) : Prop :=
  [Φ| g.pc] GSN.refines (g,l)
\end{minted}
\vspace{-2mm}
The above definition type-checks since Lean knows that \token{vGoal} is a variational extension of \token{Goal} and that \token{vGSN} is a variational extension of \token{GSN}. We can then recursively extend this definition to entire \token{vGSN} ACs to define a variational analog (\token{vGSN.deductive}) of the \token{GSN} predicate \token{deductive}. We can verify that this analog  is a correct ``lift'' of the original predicate:
\vspace{-2mm}
\begin{minted}[fontsize=\footnotesize]{lean}
theorem deductive_of_varProof {A : vGSN Φ} :
  vGSN.deductive ↔ [Φ| A.root.pc] GSN.deductive A
\end{minted}
\vspace{-0.1mm}
The above theorem allows us to verify deductive correctness for all products (i.e., the variational proof on the RHS) to producing variational proofs of \token{refines} for each variational strategy individually (i.e., via valid templates).

\textbf{Lifting Templates.} We now turn to the problem of lifting templates, i.e., step C3 of \process (Fig.~\ref{fig:workflows})
{We begin by defining the type \token{vTemplate} of variational decomposition templates. As per the semantics of $\token{vGSN}$, variational templates are defined by product-level predicates, but are instantiated on SPL-level data (i.e., variational types).
The notion of validity for variational templates is defined analogously to that of product-level templates: a variational template is valid iff every instantiation of the template (with SPL-level data) produces a sound argument. We consider the general case of lifting a conditionally valid product-level template $T$ with instantiation function ${\tt T.inst}:A \times D \to \token{List}~\token{GSN}$ and precondition $\token{T.prec} : A \times D \to {\tt Prop}$. If we have variational extensions $\alpha$ and $\gamma$ of $A$ and $D$, respectively, the problem of lifting $T$ to a variational Template $T^\prime$ becomes well-defined: we must define an instantiation function ${\tt T^\prime.inst} : \alpha \times \gamma \to \token{List}~(\token{vGSN}~\Phi)$ such that Def.~\ref{def:lift} holds. If this lifting is correct (Def.~\ref{def:lift}), then the instantiation of the template is provably sound whenever we have a \emph{variational proof} of the precondition over the set of products under study. The correctness of this process is proved in Lean as the following theorem: 
\vspace{-2mm}
\begin{minted}[fontsize=\footnotesize]{lean}
theorem lift_sound (T:Template A D) (T': vTemplate α γ) (φ:PC Ξ) :
    valid T ∧ isLift T.inst T'.inst →
    ∀ x d, [Φ| φ] T.prec (x,d) → refines (T.parent x) (T'.inst (x, d)) 
\end{minted}
\vspace{-2mm}
where $\token{x} \in \alpha$, $\token{d} \in \gamma$ and $\phi$ is the presence condition over which the template is instantiated. Note that the variational parent goal of the instantiation uses the same parent predicate as the original template $\token{T}$, applied to variational data \token{x}. Intuitively, the above theorem provides a structured approach to ``lifting'' the correctness of product-level templates, which is established during step C2 of \process. We thus reduce lifting and verifying PL AC templates to (i) lifting instantiation functions and (ii) creating variational proofs. 
\vspace{-0.1in}
\begin{example}[Variational Domain Decomposition] {We can lift the domain decomposition template~\cite{viger2023foremost} to the SPL-level and prove the correctness of the lift. In the product setting (i.e., during stage C2), we formalized a predicate \token{complete} as the precondition for validity of domain decompositions. We now need to lift the instantiation of domain decompositions to SPL-level data. We define the variational types ${\tt vSet}$ and ${\tt vFamily}$ of variational sets and families of sets, respectively. Variational sets are sets of elements annotated with presence conditions~\cite{shahin2021towards}, and variational families are finite sets of annotated sets, i.e., $\{(X_1,\phi_1),...(X_n,\phi_n)\}$. It is straightforward to lift the instantiation of domain decomposition, by setting the presence condition of each (variational) subgoal $g_i$ to be the presence condition $\phi_i$. Once we prove the correctness of the lift, we can prove the validity theorem for variational domain decomposition:
\vspace{-2mm}
\begin{minted}[fontsize=\footnotesize]{lean}
theorem vDomDecompInstLift : isLift DomainDecomp.inst vDomainDecomp.inst
theorem vDomDecompValid (S : vSet α) (F : vFamily α): 
[Φ] complete (S,F) → refines DomainDecomp.parent S (vDomDecomp.inst (S, F)) 
\end{minted}

\vspace{-2mm}
{To produce the needed variational proofs, we can either (i) lift a verifier for product-level completeness, or (ii) define fixed decompositions are completene \emph{a priori}. We give two simple examples of the latter for {finite} variational sets (\token{vFinset}). The first is a function ${\tt explode}$ which takes a finite set $S = \{(x_1,\phi_1),...(x_n,\phi_n))\}$ of annotated elements, and returns the family of annotated singletons $\{(\{x_1\},\phi_1),...(\{x_n\},\phi_n)\}$.  The second is a function ${\tt aggregate}$ which forms a variational family$\{(X_1,\phi_1),...(X_k,\phi_k)\}$}, such that elements of each $X_i$ share the same presence condition $\phi_i$. We have given reference implementations in Lean for both approaches, and provided variational correctness proofs:}

\begin{minted}[fontsize=\footnotesize]{lean}
theorem explode_complete {S : vFinset α} : [Φ] complete (explode S)
theorem aggregate_complete {S : vFinset α} : [Φ] complete (aggregate S)
\end{minted}
\end{example}
\vspace{-0.1in}}
{\bf Arguing over Lifted Analyses.}
As described in Sec.~\ref{sec:process}, \process uses both lifted templates and lifted software analyses to drive AC development. The lifting of analyses is well-understood from the SPLE literature, and we have formalized the lifting of templates above. But analysis and argumentation are not always decoupled. Consider, for instance, the argument over model checking shown in Fig.~\ref{fig:gsn-example}. The semantics of this argument -- and therefore its correctness -- is defined against a specific analysis (model checking). In following \process, we would like to replace this product-level analysis with a lifted one, without compromising argument soundness. Our formalization makes this process -- and its correctness proof -- quite straightforward.

We first formalize a general class of argument templates, which we refer to as \emph{analytic templates}, each of which is defined with respect to a fixed analysis $f$. Intuitively, an analytic argument makes assertions about the input to the analysis $x$, the output of the analysis $f(x)$, and the analysis $f$ itself. 
In general, the input of the analysis should have some relation to the subject of the goal being decomposed. For instance, the subject of \token{G0} in Fig.~\ref{fig:gsn-example} can be interpreted as a pair $(B,P)$, where $B$ is the set of behaviours executed by the (real) system, and $P$ is the set of behaviours characterized by the natural language specification. Doing model checking with model $\token{M\_SYS}$ and formal specification $\token{ALARM\_SPEC}$ is only useful if these are correct formalizations of $B$ and $P$, respectively -- hence the subgoals \token{G1} and \token{G2}. We formalize the type \token{AnalyticTemplate} in Lean following this general structure. Suppose now that we have defined an analytic template $T$ for analysis $f$, proven its validity as part of stage C2, and implemented a lifted analysis $F$. We define a general construction for {lifting} $T$ to a variational analytic template which is defined in terms of $F$ rather than $f$. Thanks to the variational semantics of \token{vGSN}, we don't need to redefine any of the predicates used in the original templates. Beyond the substitution of $F$ for $f$, the only modification is that (i) the lifted template is instantiated with variational data, and (ii) we include a claim asserting that $F$ is a correct lift of $f$. It follows that the lifted analytic template inherits the validity of the original template:

 \vspace{-3mm}
\begin{minted}[fontsize=\footnotesize]{lean}
theorem liftedAnalyticValid (T : AnalyticTemplate) 
    (F : α → β) : T.valid → [Φ] valid (T.lift F)
\end{minted}
\vspace{-3mm}

We also prove an analogous theorem for arguments over quasi-lifted analyses. Once again, our construction shows how correctness evidence produced for templates at the product level (stage C2) can be systematically lifted to the SPL-level. 

In this section, we have provided rigorous foundations for \process. Our Lean formalization of variational types was used to implement a generic framework for studying PL ACs, variational templates, and their correctness. In particular, we have shown that the process of lifting and verifying arguments for PL ACs can effectively be reduced to lifting template instantiation functions and creating variational proofs, and shown that for the general class of analytic templates, product-level analyses can be substituted for lifted analyses while preserving argument validity.

\vspace{-0.1in}
\section{Tool-Supported Product Line Assurance Case Development}
\label{sec:tool}

\label{sec:toolCaseStudy}
\vspace{-0.1in}

Any integration of formal methods to AC development should include extensive tool support, as most AC developers are not formal methods experts. {To this end, we have developed tool support for lifted AC development as part of an Eclipse-based model management framework. Our tool aims to support the \process methodology and closely follows the formalization outlined in Sec.~\ref{sec:gsnSPL}. {In this section, we  provide a brief overview of our tool (Sec.~\ref{sec:tool}), and demonstrate the feasibility of \process and our tooling on a small case study (Sec.~\ref{sec:casestudy}).}}

\vspace{-0.15in}

\subsection{A Model Management Tool for Product Line ACs}
\vspace{-0.15in}
\emph{MMINT}~\footnote{\url{https://github.com/adisandro/MMINT}} is a Eclipse-based model management framework developed at the University of Toronto. It is a generic framework which can be extended with plugins for specific modeling tasks. One of its extensions, \toolName~\cite{minta}, is used for model-driven AC development, supporting GSN modeling and model-based analyses. Another extension, \toolNamePL, supports product line modelling and lifted model-based analyses~\cite{di2023adding}.

Several functionalities of \toolName\; and \toolNamePL\; can be reused directly to support \process. We combine \toolName's GSN metamodel~\cite{minta}, and
\toolNamePL's generic variational metamodel (GVM,~\cite{di2023adding}) to define a metamodel for product lines of GSN ACs. To support \process, we also needed to implement some new functionalities: (i) we extended \toolName's GSN template module to recognize product line models, such that instantiation of GSN templates can be done with either product-level or lifted versions; (ii) users can define product-level analytic templates as formalized in Sec.~\ref{sec:gsnSPL}, allowing the results of specified analyses to be weaved into ACs as part of instantiation. When these templates are instantiated on variational models, if the analyses associated with the template have been lifted, the lifted analyses are executed, and the lifted template is instantiated instead; (iii):  \toolNamePL's GVM is unable to provide an appropriate visualizations for arbitrary product line models, so we created a custom visualization module to facilitate the manual inspection of PL ACs.

\vspace{-0.15in}
\subsection{Case Study: Assuring a Product Line of Infusion Pumps with \process}
\label{sec:casestudy}
\vspace{-0.15in}
{To demonstrate the feasibility of the \process methodology and the features of our AC development tool, we followed our proposed methodology to create a partial AC for a product line of medical infusion pumps.}
\vspace{-0.2in}
\paragraph{System Details.}
Infusion pumps are devices used to administer medication or other fluids to patients. As different patients may have different medical needs, it is natural to model a SPL of infusion pumps with different optional features.
{We began from an existing Extended Finite State Machine (EFSM) model of an infusion pump created as part of a multi-institute research project~\cite{alur2004formal}. While this model was not originally defined as a  SPL, it was designed to model features and hazards for an infusion pump \emph{in general}, the authors noting that in general ``no single device [...] has all of the design features''~\cite{zhang2010hazard}. We extended the EFSM to a SPL by mapping six optional features to their associated states and transitions and annotating these elements with presence conditions. For example,}
{\token{CHECK\_INFUSION\_RATE} is an optional feature that allows a pump to monitor the current rate of delivery of a drug.} 
{The resulting SPL encompasses a family of 36 valid product configurations.  
{A more detailed description of the SPL model is given in Appendix~\ref{app:caseStudy}.}}  In our hypothetical assurance scenario, the top-level assurance obligation (i.e., the root node of the AC) is to show that when an alarm is triggered (e.g., due to a dosage limit violation), the system will not administer a dose until the alarm is disabled.

\paragraph{Step C1: Defining the Assurance Process.}
Before beginning AC development, we need to determine which types of analyses and argumentation are applicable for our assurance task. For simplicity, we considered two kinds of analyses: querying of models (via the Viatra Query Language~\cite{vql}) and model checking~\cite{baier2008principles}. As part of the assurance process, we require that the use of these analyses be accompanied by sufficient assurance that the models and specifications used for analyses have been validated. We also allow for the use of domain decomposition templates~\cite{viger2020just} to break down assurance obligations.
\vspace{-3mm}

\paragraph{Step C2: Formalization.}
Based on the assurance process defined in Step 1, we can now formalize the associated argument structures. The assurance obligations associated with using queries and model checking can be formalized as analytic templates, analogous to the argument shown in Fig.~\ref{fig:gsn-example}. We can do a (shallow) formalization of the semantics of both analyses in Lean, and verify both analytic templates. The domain decomposition template has been formalized as described in Sec.~\ref{sec:gsnSPL}.

\vspace{-3mm}
\paragraph{Step C3: Lifting.}
We now lift the analyses and templates formalized in Step 2 so that they can be applied directly to SPLs. For lifted model queries, we reused a lifted query engine developed as part of previous work~\cite{di2023adding}. For (quasi-)lifted model checking, we used the tool FTS2VMC~\cite{ter2022efficient} which verifies featured transition systems (FTSs) against specifications written in an action-based branching time logic (v-ACTL). The templates formalized for both analyses can be lifted automatically via the constructions for lifted and quasi-lifted analytic templates (Sec.~\ref{sec:gsnSPL}). We can also employ the lifted constructions for domain decompositions given in Sec.~\ref{sec:gsnSPL}. 

\paragraph{Step C4: Lifted AC Development}

\begin{figure}[t]
    \centering
\includegraphics[width=0.8\textwidth]{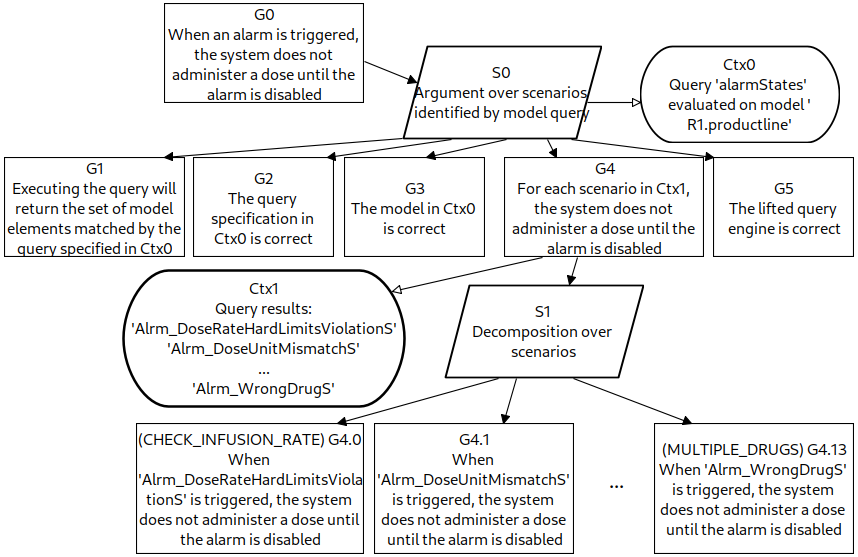}
\vspace{-0.1in}
    \caption{Instantiations of the lifted query and scenario decomposition templates.}
    \label{fig:query_template}
\end{figure}

Having defined, formalized and lifted the analyses and templates for our task, we can perform lifted AC development. We begin with the root goal asserting the primary safety property (not administering drug doses while an alarm is active). Under normal circumstances, we could proceed by running a model query which will return all alarm states in the model, and then assure the property for each alarm scenario. Thanks to our the lifting of the query engine, we can apply the same rationale using our lifted analyses and argument templates, even though different products may be associated with different sets of alarm states. Fig.~\ref{fig:query_template} shows the results of instantiating the lifted query template followed by a lifted domain decomposition. When we instantiate the generic model query template, the tool automatically detects that the model is an SPL, and executes the lifted query engine to return a variability-aware set of query results (\token{Ctx0}). These results are then woven into the AC using a lifted version of the analytic template. This lifted analytic argument instantiation is sound as per theorem \token{liftedAnalyticValid} (Sec.~\ref{sec:gsnSPL}) and our original proof that the product-level template was deductive.  {Returning to the AC, we proceed from goal \token{G4} using a variational domain decomposition, assigning each alarm state to its own goal, such that the presence conditions identified by the lifted query result (e.g., \token{CHECK\_INFUSION\_RATE}) are used to annotate each subgoal}. {The absence of a presence condition (e.g., \token{G3.1}) means that the alarm is present in every product.} This decomposition corresponds to the verified \token{explode} construction and is thus also sound by theorem \token{vDomDecompValid} (Sec.~\ref{sec:gsnSPL}).

We can then continue to produce assurance for each identified alarm scenario in a lifted fashion. We focus on goal \token{G4.0}, which effectively asserts that every product with feature \token{CHECK\_INFUSION\_RATE} satisfies the given safety property in the context of alarms due to dose rate violations. This can be verified using (lifted) software model checking. As with queries, we can instantiate the model checking template formalized in Step 2 and lifted in Step 3, using the quasi-lifted model checker FTS2VMC for family-level verification. We formalize the property in goal \token{G4.0} as 
\textbf{AG} ({\token{Alrm\_DoseRateHardLimitsViolationS}} $\Rightarrow$ \textbf{A}[!(\token{Infusion\_NormalOperationS}) \textbf{U} (\token{E\_ClearAlarm})])
and run the lifted model checker on the infusion pump FTS model. In this case, the model checker does not reveal any violations, meaning that every product with \token{CHECK\_INFUSION\_RATE} satisfies the given property. This can then be incorporated as variational evidence for this family of products, as shown in Fig.~\ref{fig:model_check_ac}.  {The remaining pieces of evidence (e.g., \token{G7}, that the formalization is correct) still need to be produced. Note that in \token{G10}, the correctness of the lift is interpreted as correctness of {quasi}-lifting. Furthermore, evidence for \token{G6} and \token{G10} needs to be produced {only once} and can be reused in subsequent applications of the template.} Once we have provided the required evidence for this argument (and the analogous evidence for the lifted query), we can then repeat this verification process for each alarm scenario until all assurance obligations have been satisfied.

\begin{figure}[t]
    \centering
\includegraphics[width=0.8\textwidth]{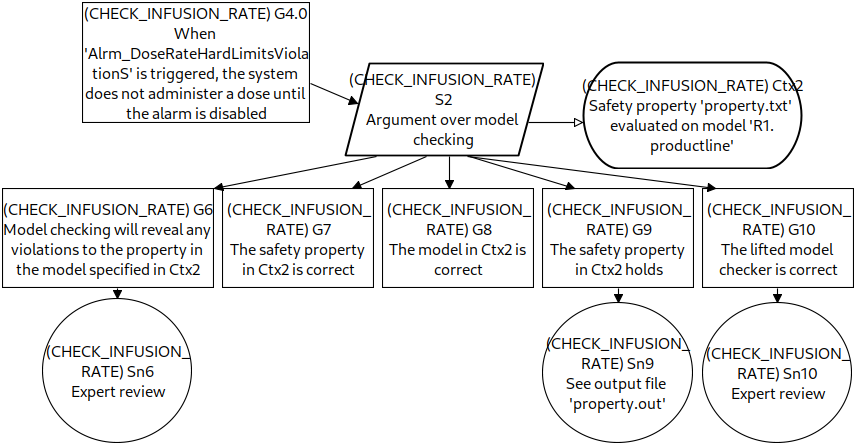}
\vspace{-0.1in}
    \caption{Tool-generated lifted analytic model checking template instantiation for claim \token{G4.0}.}
    \label{fig:model_check_ac}
    \vspace{-0.25in}
\end{figure}

\noindent\textbf{Observations.} {In developing the above AC fragment for the infusion pump SPL, we demonstrated the feasibility of \process to support multi-layered AC development in a lifted fashion. We emphasize two specific points. (1) {By following \process, {the {operational process} of AC development becomes essentially the same as in product-level AC development}}. That is to say, it suffices to know what one would do to assure a single product, and the lifted analyses and templates can correctly generalize this knowledge to the SPL-level. This is due to the particular variational semantics adopted by \process, as outlined in Sec.~\ref{sec:gsnSPL}. (2) Traditionally, the primary use of software analyses in AC development has been to produce evidence artifacts. Lifted analyses can correspondingly produce variational evidence. But they can also be used to \emph{systematically identify {variation points} in the SPL which are relevant to the design of the AC}. For example, in our case study, the variation points identified by the lifted query led systematically to \emph{structural variability} in the AC (i.e., the subgoals of strategy \token{S1} in Fig.~\ref{fig:query_template}).}

\noindent\textbf{Limitations and Threats to Validity.} One of the limitations of the tooling is that modeling languages and model-based analyses must be defined natively in order to leverage the GVM. {For instance, the lifted model checker is not a natively defined model-based analysis, and its verification results are not automatically interpreted as variational data (this contrasts with querying, which is a native model-based analysis)}. {There is also no automated support for template verification; all argument templates are verified manually ahead of time. Part of our future work is integrating the modelling tool with our Lean formalization to provide theorem proving support during AC development.}

With respect to the validity of our observations, we note that our partial AC is relatively narrow in scope and was designed by the authors for the purpose of demonstrating \process. While we believe we have successfully demonstrated the feasibility of our methodology, further empirical validation is required. Ideally, this evaluation can be done as part of a collaboration with industrial assurance engineers, since real-world assurance processes may be more difficult to formalize and lift than those shown here.
\vspace{-3mm}
\section{Related Work}
\label{sec:related}
\paragraph{Analysis of Software Product Lines.}
Implementing scalable analyses of software product lines is one of the central problems in SPLE. Thum et al.~\cite{thum2014classification} divide SPL-level analysis strategies as either \emph{product-based} (e.g., brute-force or sampling-based techniques), \emph{feature-based} (i.e., analyzing feature modules independently and composing the results), or \emph{family-based} (i.e., operating at the level of sets of products). Lifted analyses are a specific form of SPL-level analyses, which can be implemented using any of these three strategies. A wide variety of analyses have been lifted in the SPLE literature, e.g.~\cite{classen2014formal,thum2012family,kastner2012type,salay2014lifting,shahin2019lifting}. We have previously proposed a catalogue of \emph{lifting techniques}, i.e., approaches to implementing lifts with various tradeoffs in terms of engineering and validation effort. Our formalization of PL AC semantics in Lean is preceded by various formalizations of SPL analyses and reasoning frameworks in proof assistants. Lifted analyses which have been formally verified include type checking~\cite{kastner2012type} and model-based change impact analyses~\cite{shahin2021towards}. Using a similar formalization of configuration semantics, Alves et al. formalized a theory of SPL refinement in Coq~\cite{alves2020porting}. Their formalization uses a lower-level definition of SPLs as compared to variational types, requiring users to explicitly define a particular form of mapping from features to software assets in order to define an SPL instance. By contrast, our notion of variational types is purely algebraic, only requiring users to define a suitable derivation operator from the variational type to a product type.

\paragraph{Assurance Cases for Product Lines.}
{{Assurance cases for SPLs were studied by Habli and Kelly~\cite{habli2009model,habli2010safety}, who argued that safety-relevant variation points in the SPL should be reflected explicitly in the product line AC. Habli and Kelly propose using GSN patterns and the modular GSN extension~\cite{habli2010safety} to represent product line ACs. Habli~\cite{habli2009model} also provides an SPL safety metamodel which allows for variation points in the SPL to be traced explicitly to a PL AC. This approach was further refined in de Oliveira et al.~\cite{de2015supporting}, introducing tool-supported generation of modular PL ACs from feature-based system models and safety analyses. However, this approach does not provide a distinction between product-level and SPL-level semantics or analysis, implicitly assuming that analysis and reasoning are performed at the product. As an alternative to product-based AC development, Habli~\cite{habli2009model} also considers a primarily {feature-based} AC development method, in which AC modules are developed independently for each feature, and then the  PL AC is obtained by composing these modules. However, this method requires the assurance engineer to either (a) identify and mitigate all potential feature interactions, which can devolves into to the brute-forced assurance engineering, or (b) tolerate an incomplete assessment of potential feature interactions. By contrast, \process supports analysis and reasoning over {all} valid configurations of the SPL, without resorting to product-level work. To the best of our knowledge, the only existing variability-aware AC development process was proposed by Nešić et. al~\cite{nevsic2021product}, which lifts contract-based templates to PLs of component-based systems~\cite{nevsic2021product}. \process is effectively an attempt to generalize the approach of Nešić et. al to {arbitrary} templates and analyses.}

\paragraph{Formal Methods for Assurance Case Development.}
The most obvious use of formal methods for AC development is for the production of evidence; well-established verification techniques such as model checking~\cite{baier2008principles} and deductive verification~\cite{leino2010dafny} provide invaluable evidence for formally specified requirements. As we have mentioned above, the integration of formal methods with AC development requires extensive tool support. The AC development tool AdvoCATE~\cite{Denney:2018} uses the AutoCert inference engine~\cite{denney2014automating} to check system software implementations against formal specifications, which are then grafted into the AC following a predefined template. The Evidential Tool Bus (ETB)~\cite{Cruanes:2013} gathers verification evidence from various external tools, and then creates ACs from the bottom up using compositional rules written in a variant of Datalog.} Formal methods can also be used to ensure that instantiations of argument templates actually yield sound arguments. 
{As ACs and proofs are closely related, proof assistants have been used on several occasions to study rigorous AC development. Rushby~\cite{rushby2009formalism} demonstrated a proof-of-concept of how an AC could be modelled using the proof assistant PVS. The assurance case editor D-Case was extended with a translation to the Agda programming language, such that an AC could be specified as an Agda program to check for well-formedness~\cite{takeyama2011brief}. An extension to Isabelle was developed to embed the Structured Assurance Case Metamodel (SACM) as part of its documentation layer~\cite{foster2021integration}. Finally, Viger et al.~\cite{viger2023foremost} used Lean to study the correctness of decomposition templates for model-based assurance cases. To the best of our knowledge, our formalization of ACs in Lean is the first to support the formalization of entire ACs at the using nested inductive types, rather than only individual claims~\cite{foster2021integration} or individual strategies~\cite{viger2023foremost}.}

\vspace{-0.1in}
\section{Conclusion}
\vspace{-0.1in}
\label{sec:conclusion}
{In this work, we proposed \process, an assurance engineering methodology supporting lifted assurance case development for software product lines. \process extends existing formal approaches to AC development by lifting formal argument templates and software analyses to the SPL-level. We used the proof assistant Lean to formalize rigorous foundations for \process. We demonstrated the feasibility and usefulness of our methodology by developing a partial PL AC over a product line of medical devices. 

{We identify several avenues for future work. One is to further extend our model-based development framework for \process with support for additional lifted templates and analyses, and to integrate the modeling layer with the formalization layer, which would enable automated or semi-automated theorem proving to assist in AC development~\cite{viger2023foremost}. Another interesting line of work is the problem of efficiently creating and maintaining variational proofs (Sec.~\ref{sec:gsnSPL}). While such proofs can be constructed in an \emph{ad hoc} fashion, a systematic approach for defining, certifying, and repairing such objects would be invaluable for proof-driven PL AC development and maintenance.
{In particular, language-independent techniques for maintaining and repairing variational proofs over evolutions of SPLs could be leveraged to provide assurance engineers with formal guarantees about the maintenance of SPL-level assurance.}

\bibliographystyle{splncs04}\bibliography{main}
\appendix
\setminted{baselinestretch=1}
\section{Notes on our Formalization}
\label{app:formal}
Many of the Lean code snippets presented in this paper (including this appendix) are slightly modified for ease of presentation and legibility. We often omit implicit parameters, termination proofs, auxiliary typeclass instances, and function decorators. The complete formalization is available at {\url{https://github.com/loganrjmurphy/placs}}.

\subsection{Formalizing Variability Semantics}
We allow any type with decidable equality and finitely many elements to be used as a language of features. Thus, we define \token{FeatureSet} as a typeclass: 
\begin{minted}[fontsize=\footnotesize]{lean}
class FeatureSet (α: Type) extends Fintype α where
  dec: DecidableEq α

-- canonical example : finite sets of natural numbers 
instance {n : Nat} : FeatureSet (Fin n) := .mk inferInstance
\end{minted}
Feature expressions are defines as propositional expressions over a set of features as follows:
\begin{minted}[fontsize=\footnotesize]{lean}
inductive FeatExpr (Ξ : Type) [FeatureSet Ξ] : Type
| all  : FeatExpr Ξ
| none : FeatExpr Ξ
| atom : Ξ → FeatExpr Ξ
| not  : FeatExpr Ξ → FeatExpr Ξ
| and  : FeatExpr Ξ → FeatExpr Ξ → FeatExpr Ξ
| or   : FeatExpr Ξ → FeatExpr Ξ → FeatExpr Ξ
deriving DecidableEq

\end{minted}
Feature models and presence conditions are just type aliases for feature expressions.
\begin{minted}[fontsize=\footnotesize]{lean}

@[reducible]
def FeatModel (Ξ : Type) [FeatureSet Ξ] := FeatExpr Ξ

@[reducible]
def PC (Ξ : Type) [FeatureSet Ξ] := FeatExpr Ξ

\end{minted}

Next, we define a space of configurations (\token{ConfigSpace}) as finite sets of configurations (confiugrations themselves being finite sets of features). Since we often want to refer to configurations with respect to some feature model, we actually define the semantics of feature expressions in terms of \token{ConfigSpace}, and only then define configurations themselves.

\begin{minted}[fontsize=\footnotesize]{lean}
def semantics {Ξ: Type} [Ξ]: FeatExpr Ξ → ConfigSpace Ξ
| .top => Finset.univ
| .bot => ∅
| .atom f => Finset.univ.filter (f ∈ .)
| .not FeatExpr => (semantics FeatExpr).compl
| .and pc1 pc2 => (semantics pc1) ∩ (semantics pc2)
| .or pc1 pc2 => (semantics pc1) ∪ (semantics pc2)

def Conf  (Φ : FeatModel Ξ) : Type := { c : Finset F // c ∈ semantics Φ}

def sat  (c : Conf Φ) (p : FeatExpr Ξ) : Prop := c.val ∈ semantics p

infix:50 "⊨" => sat

\end{minted}
\subsection{Formalizing Goal Structuring Notation (GSN)}
For the sake of simplicity, our GSN formalization focuses on goals and evidence, since these are the only components with essential semantic content. Strategy nodes, while essential for readability of ACs, do not introduce any new assurance obligations. We define a {goal} as either an atomic proposition, or the application of some predicate $P$ to an object $x$. Note that the predicative constructor is polymorphic, so we can use predicates over any arbitrary datatype. 
\vspace{-2mm}
\begin{minted}[fontsize=\footnotesize]{lean}
inductive Goal
| atom (p : Prop) : Goal 
| pred {α : Type} (P : α → Prop) (x : α) : Goal
\end{minted}
\vspace{-2mm}
We then define the language of GSN trees -- our formal language of assurance cases -- via three constructors. Our formalization is essentially the same as the GSN grammar described by Matsuno et als~\cite{matsuno2014design}. The empty constructor \token{nil} represents an empty GSN tree. The evidence constructor \token{evd} allows to produce evidence for any goal $g$ by supplying a proof of the proposition $\llbracket g \rrbracket$ denoted by $g$. The inductive constructor \token{strategy} allows us to decompose any goal into finitely many subtrees (its children). A goal is considered ``undeveloped'' if either its children are the empty list, or a list containing only \token{nil} nodes. 
\vspace{-2mm}
\begin{minted}[mathescape=true,fontsize=\footnotesize]{lean}
inductive GSN
| nil : GSN    
| evd (g : Goal) : ⟦g⟧ → GSN
| strategy : Goal → List GSN → GSN
\end{minted}

\subsection{Formalizing Variational GSN}
\label{app:vGSN}
To formalize \token{vGSN}, we begin with the type \token{vGoal} of variational goals. As stated in   Sec.~\ref{sec:gsnSPL}, the semantics of variational goals are provided by the combination of product-level predicates with SPL-level data. We also allow for \token{vGoals} to consist of atomic non-variational propositions.
\begin{minted}[mathescape=true,fontsize=\footnotesize]{lean}
inductive vGoal (Φ : FeatModel F)
| atom (pc : PC F) (p : Prop) 
| pred {A α : Type} [Var α A Φ] (pc : PC F)  (P : A → Prop) (x : α)
\end{minted}

The definition of \token{vGSN} shown in Sec.~\ref{sec:gsnSPL} was modified to emphasize the role of variational proofs as evidence for variational predicative goals. Since our definition of ${\tt vGoal}$ also supports non-predicative goals, the proper definition of \token{vGSN} is as follows:

\begin{minted}[mathescape=true,fontsize=\footnotesize]{lean}
inductive vGSN (Φ: FeatModel F)
| evd (g : vGoal Φ) : g.varProof → vGSN Φ
| strategy : vGoal Φ → List (vGSN Φ) → vGSN Φ
\end{minted}
where ${\tt g.varProof}$ reduces to a variational proof if $g$ is a predicative goal, and a normal proof it is not. Note that ${\tt vGSN}$ (and indeed ${\tt GSN}$) are nested inductive types, meaning they are not structurally recursive. This means that we occasionally need to provide termination proofs for functions on ACs (and PL ACs), and we need to provide an explicit induction principle. Thankfully, we can prove the natural induction principle for ${\tt vGSN}$ as a theorem: 

\begin{minted}[mathescape=true,fontsize=\footnotesize]{lean}
@[induction_eliminator]
def vGSN.inductionOn
  (motive : vGSN Φ → Prop)
  (evd : ∀ (g : vGoal Φ) (e : g.varProof), motive (vGSN.evd g e))
  (strategy :
     ∀ (g : vGoal Φ) (children : List (vGSN Φ)),
         (∀ (g : vGSN Φ), g ∈ children → motive g)
            → motive (strategy g children))
  : ∀ (G : vGSN Φ), motive G 
\end{minted}

An example of a function requiring a termination proof is the derivation operator from ${\tt vGSN}$ to ${\tt GSN}$:

\begin{minted}[mathescape=true,fontsize=\footnotesize]{lean}
def derive (c : Conf Φ) : vGSN Φ → GSN
| .evd g e =>
  if h:c ⊨ g.pc then .evd (g ↓ c) (e c h) else .nil
| strategy g children=>
  if c ⊨ g.pc then
    GSN.strategy (g ↓ c) (mapFilterNil children (derive c))
  else .nil
decreasing_by
  simp_wf
  cases x <;> (rename_i h; have := List.sizeOf_lt_of_mem h; omega)    
\end{minted}
where ${\tt mapFilterNil}$ maps the derivation operator over the list of children and removes those which become ${\tt nil}$.

\section{Case Study Details}
\label{app:caseStudy}

Our case study uses a feature transition system model to represent a family of medical infusion pumps with different configurations of optional features. To create our model, we began with an extended finite state machine model from the literature that represents an abstract, generic infusion pump~\cite{alur2004formal}. This model represents both the core functionalities of an infusion pump (i.e., delivering insulin doses to patients at a predefined safe rate) and a number of optional functionalities that are not necessary for the pump to perform its main function, such as the ability to administer multiple different types of drugs. Then, using a list of possible features for infusion pumps and their corresponding hazards from the literature~\cite{zhang2011generic}, we identified a number of model elements from the original EFSM that only exist if certain optional features are present. We converted the EFSM into a variability-aware feature transition system by annotating each model element with presence conditions corresponding to its required features. The following six features were used to annotate the model:

\begin{itemize}
    \item \texttt{HW\_MONITORING} enables the pump to monitor its hardware components, raising alarms when hardware errors are detected.
    \item \texttt{MULTIPLE\_DRUGS} indicates that the pump is designed to administer multiple different types of drugs other than just insulin.
    \item \texttt{CHECK\_DRUG\_TYPE} allows the pump to detect which drug is ready to be administered, and is a necessary safety feature if the pump is used to administer multiple different drug types.
    \item \texttt{PROGRAMMABLE\_INFUSION} allows the concentration and dosage rate of infusions to be customized, whereas pumps without this feature must stick to a specific predefined infusion rate.
    \item \texttt{CHECK\_INFUSION\_RATE} enables the pump to detect the rate that an infusion occurs at, and is a necessary safety feature if the \texttt{PROGRAMMABLE\_INFUSION} feature is present.
    \item \texttt{VISUAL\_DISPLAY} provides a visual interface to display details about a given infusion.

\texttt{CHECK\_DRUG\_TYPE} is required when \texttt{MULTIPLE\_DRUGS} is present in order to ensure that an incorrect drug is never administered, and \texttt{CHECK\_INFUSION\_RATE} is required when \texttt{PROGRAMMABLE\_INFUSION} is present in order to ensure that users do not set the infusion rate to unsafe thresholds. The feature model for the infusion pump product line is therefore given as follows:

(\texttt{MULTIPLE\_DRUGS} => \texttt{CHECK\_DRUG\_TYPE}) \& \\(\texttt{PROGRAMMABLE\_INFUSION} => \texttt{CHECK\_INFUSION\_RATE})
\end{itemize}

Of the $2^{6} = 64$ distinct combinations of these features, 36 configurations satisfy the feature model.
\end{document}